\documentclass[11pt]{article}

\usepackage[dvips]{graphics}
\usepackage{amsmath}
\usepackage{amsfonts}
\usepackage{amssymb}
\usepackage{latexsym}
\usepackage{graphics}
\usepackage{cite}
\usepackage{lscape}
\usepackage{epsfig}
\usepackage{color}

\renewcommand{\d}{\mathrm{d}}

\newcommand{\pp}[2][]{\ensuremath{\frac{\partial #1}{\partial #2}}}

\newcommand{\captn}[1]{\vspace{-3ex}\caption{\small #1}}

\DeclareMathSymbol{\mg}{\mathrel}{symbols}{"1D}

\newcommand{\ga}{\alpha}
\newcommand{\gb}{\beta}
\renewcommand{\gg}{\gamma}
\newcommand{\gd}{\delta}
\renewcommand{\ge}{\epsilon}
\newcommand{\gve}{\varepsilon}
\newcommand{\gf}{\phi}

\newcommand{\gx}{\xi}
\newcommand{\gm}{\mu}
\newcommand{\gn}{\nu}

\newcommand{\gr}{\rho}
\newcommand{\gth}{\theta}

\newcommand{\gs}{\sigma}

\newcommand{\gp}{\pi}
\newcommand{\gps}{\psi}
\newcommand{\get}{\eta}

\newcommand{\gD}{\Delta}
\newcommand{\gF}{\Phi}

\newcommand{\gL}{\Lambda}

\newcommand{\gTh}{\Theta}

\newcommand{\gP}{\Pi}

\newcommand{\te}{{\tilde e}}

\newcommand{\tg}{{\tilde g}}

\newcommand{\Id}{\text{\small 1}\hspace{-3.5pt}\text{1}}

\newcommand{\ra}{\rightarrow}

\newcommand{\der}{\partial}

\newcommand{\inv}{^{-1}}
\newcommand{\dsp}{\displaystyle}

\newcommand{\beq}{\begin{equation}}
\newcommand{\eeq}{\end{equation}}
\newcommand{\barr}{\begin{array}}
\newcommand{\earr}{\end{array}}
\newcommand{\equ}[1]{\begin{gather} #1 \end{gather}}
\newcommand{\equa}[1]{\begin{align} #1 \end{align}}

\newcommand{\tabu}[2]{\begin{tabular}{#1} #2 \end{tabular}}
\newcommand{\arry}[2]{\begin{array}{#1} #2 \end{array}}

\newcommand{\pmtrx}[1]{\begin{pmatrix} #1 \end{pmatrix}}

\newcommand{\non}{\nonumber}
\newcounter{oldcounter}

\newcommand{\bL}{{\bar L}}

\newcommand{\ba}[2]{\[\begin{array}{#2}\label{#1}}
\newcommand{\ea}{\end{array}\]}
\newcommand{\be}{\begin{equation}}
\newcommand{\ee}{\end{equation}}
\newcommand{\bea}{\begin{eqnarray}}
\newcommand{\eea}{\end{eqnarray}}

\begin{document}

\thispagestyle{empty}

\begin{flushright}
UMN-TH-2523/06 \\ 
HD-THEP-06-26 \\ 
SIAS-CMTP-06-7\\
hep-th/0610169
\end{flushright}
\vskip 2 cm
\begin{center}
{\Large {\bf
Nonlinear Properties of Vielbein Massive Gravity
}
}
\\[0pt]
\vspace{1.23cm}
{\large
{{\bf Stefan Groot Nibbelink$^{a,}$\footnote{
{{ {\ {\ {\ E-mail: grootnib@thphys.uni-heidelberg.de}}}}}}},
{\bf Marco Peloso$^{b,}$\footnote{
{{ {\ {\ {\ E-mail: peloso@physics.umn.edu}}}}}}},
{\bf Matthew Sexton$^{b,}$\footnote{
{{ {\ {\ {\ E-mail: sexton@physics.umn.edu}}}}}}},
\bigskip }\\[0pt]
\vspace{0.23cm}
${}^a$ {\it
Institut f\"ur Theoretische Physik, Universit\"at Heidelberg, Philosophenweg 16 und 19, D-69120 Heidelberg, Germany
 \\[1ex]
Center for Mathematics and Theoretical Physics, 
Shanghai Institute for Advanced Study,
University of Science and Technology of China,  
99 Xiupu Rd, Shanghai 201315, China
}
\\[3ex]
${}^b$ {\it
School of Physics \& Astronomy,
University of Minnesota, 116 Church Street S.E., \\
Minneapolis, MN 55455, USA
\\}
}
\bigskip
\vspace{1.4cm}
\end{center}
\subsection*{\centering Abstract}

We propose a non--linear extension of the Fierz--Pauli mass
for the graviton through a functional of the vielbein and an
external Minkowski background. The functional generalizes the notion of the
measure, since it reduces to a cosmological constant
 if the external background is formally sent to zero. Such a term and the explicit external
background, emerge dynamically from a bi--gravity theory, having both
a massless and a massive graviton in its spectrum, in a specific limit in
which the massless mode decouples, while the massive one 
couples universally to matter. We investigate the massive theory using
the St\"uckelberg method and providing a 't Hooft--Feynman
gauge fixing in which the tensor, vector and scalar St\"uckelberg
fields decouple. We show that this model has the softest possible
ultraviolet behavior which can be expected from any generic (Lorentz
invariant) theory of massive gravity, namely that it becomes
strong only at the scale $\Lambda_3 = (m_g^2 M_P)^{1/3}\,$.

\newpage
\setcounter{page}{1}

\section{Introduction and Discussion}
\label{sc:intro}


Motivated by the observed accelerated expansion of the
universe~\cite{acc}, and by the theoretical difficulties in ascribing
it to a cosmological constant, there has been considerable activity in
modifications of gravity at large scales in the past years. For
instance, an accelerated expansion can be achieved in bi-gravity 
models~\cite{damour2002}, in models in which the Lorentz symmetry is
broken by the gradient of a field~\cite{ghost}, or in four dimensional
models embedded in extra dimensions, as the self--accelerating DGP
branch~\cite{dgp,deffayet2002,instabilitydgp}. Some of these proposals
have similar properties to massive gravity, which is probably the most
straightforward and best studied modification of general relativity.


At the linearized level, massive gravity is obtained by adding to the 
Einstein--Hilbert action a mass term for the metric perturbations 
$h_{\mu \nu} = g_{\mu \nu} - \get_{\mu  \nu}$. The quadratic
Lagrangian for this massive spin--two tensor field is given
by~\cite{fipa}   
\begin{equation}
L ~=~ 
 \frac{1}{4} M_P^2 \Big\{ 
h^{\mu \nu} \Big( 
h_{,\mu \nu} + \Box h_{\mu \nu}
- {h^\alpha}_{\mu , \alpha \nu}- {h^\alpha}_{\nu,\alpha \mu}
+ \eta_{\mu \nu}
{h_{\alpha\beta,}}^{\alpha\beta} - \eta_{\mu\nu}\Box h
\Big)
+ m_g^2 \Big( h^2 - h^{\mu \nu} h_{\mu \nu} \Big)
\Big\}~,
\label{fp}
\end{equation}
where $h=\eta^{\mu\nu}h_{\mu\nu}$ is the trace of the metric
perturbation. There is a very stringent experimental bound on the
graviton mass:  $m_g \leq 7\times 10^{-41}\mbox{ GeV}$ \cite{pdg},
which is  close to the inverse of the size of the observable universe.
As already observed by Fierz and Pauli (FP)~\cite{fipa} the
relative sign between the two mass terms is fixed uniquely by the
requirement of having a ghost--free Lorentz--invariant (linear)
theory.~\footnote{A richer structure of ghost--free mass terms is
  possible if one is willing to give up Lorentz invariance~\cite{rudu}.} 
The massless linear theory $m_g = 0$ can be uniquely extended beyond
quadratic order using the requirement of general covariance leading to
the familiar Einstein-Hilbert action. But
because the mass term breaks covariance, it has no unique
non-linear extension.

Covariance can be restored by introducing additional degrees of
freedom, as for instance it is done with the St\"uckelberg
method~\cite{ags}. Another approach is to introduce a second metric
into the theory~\cite{fdominance,damour2002,gnp}. When one of the two
metrics obtains a background expectation value, a mass term for the
other metric is generated. Even though such bi--gravity theories are
covariant, their completion of the Fierz--Pauli mass term is far from
unique because one can write down an infinite set of invariant
non--linear interactions between the two metrics. It is possible to
obtain more uniquely defined bi--gravity theories: 
Ref.~\cite{gnp} considers a bi-gravity model described in terms of the
vielbeins (tetrads), rather than metrics. Besides Einstein--Hilbert
actions for both sectors it includes all possible cosmological
constant like terms, that can be written down using these two
vielbeins. The model~\cite{gnp}, reviewed in appendix~\ref{sc:bigrav},
has two spin--two fields, one of which is massless, while the other
has a mass term of the FP form~(\ref{fp}). Interestingly, the model
admits a limit in which the massless mode decouples, while the massive
one couples universally to matter.

The goal of the present paper is to investigate the resulting model of
massive gravity. We would like to perform an  analysis beyond the
linearized level, and to compare our results with those obtained for
generic massive gravity theories. In particular we want to investigate
when does this massive gravity theory become strong? This question is
important because it is related to the Van Dam-Veltman-Zhakarov (vDVZ)
discontinuity~\cite{vdvz}. The propagator of a massive graviton does
not reduce to the massless one in the limit of vanishing graviton mass
$m_g \rightarrow 0$. The discontinuity between these propagators
results in a discontinuity between the perturbative interactions of
the massless and massive theories. However, precisely because the
interactions of massive gravity become strong, this does not
necessarily mean a discontinuity between the final nonperturbative
results~\cite{arkady1,arkady2,ags}.  Ref.~\cite{ags} showed that the
scale at which any non--linear completion of the FP mass term
\eqref{fp} becomes strong never  exceeds 
$\Lambda_3 = ( m_g^2 M_p )^{1/3} \,$. Any generic 
completion, which becomes strong at a smaller energy, can be improved
by adding suitable terms, to result in a  theory which becomes strong
at $\Lambda_3\,$. This procedure is in general rather involved. Quite
remarkably, we will prove that our model becomes strong precisely at
the scale $\Lambda_3 \,$, without the need of such additional terms.

The plan of this paper is the following. Section~\ref{sc:model}
describes the model of massive gravity we want to study in this
paper using the vielbein formalism. In Section~\ref{sc:stuckelberg}, the
St\"uckelberg analysis of~\cite{ags} is extended to the case of the
vielbein. In general, the vielbein formulation of gravity contains more
(non-dynamical) fields than the  standard metric formulation, which
are compensated by additional gauge symmetries (the local Lorentz
transformations). We provide the (unique) transformation which fixes
these additional degrees of freedom. In Section~\ref{sc:propagators}
we determine the  quadratic action for the graviton and the
St\"uckelberg fields. We present a gauge choice, that explicitly
decouples the scalar, vector and spin--two degrees of freedom. This
choice, which, to our knowledge, had not been provided so far in the
literature, leads to particularly simple propagators for the different
polarizations. In Section~\ref{sc:interactions} we classify the
dominant interaction terms, and we show that the scale at which the
theory becomes strong is $\gL_3$. In that Section we also compute the
tree level amplitude of the $2 \rightarrow 2$ scalar scattering,
because it can be  considered as a typical diagram used to compute the
scale at which massive gravity becomes strong. However, we show, 
by a specific choice of parameters, that this interaction can be made
to vanish at this scale, while other interactions still remain strong.
These results are summarized in the concluding section~\ref{sc:conclusions}.
In Appendix~\ref{sc:bigrav} we explain how the
model of massive gravity studied in this paper can be obtained via a
decoupling of the bi--gravity model introduced in~\cite{gnp}. 
The following appendices are more technical.
Appendix~\ref{sc:bracket} describes some useful properties
of a special product that generalizes the definition of the
determinant, which is used to write the interaction between the two initial
gravitational sectors. The Appendices~\ref{sc:pertexp} and
\ref{sc:exact} contain useful intermediate results for the
computation of the action in terms of  the St\"uckelberg fields.

\section{A massive gravity theory described in the vielbein formalism}
\label{sc:model}

In this Section we outline the model of massive
gravity that will be studied in this paper. 
The action of this model is 
\equ{
S ~=~ \frac 12 \, M_P^2 \int d^4 x \Big\{
\sqrt{-g} \, R \,+\, 3 \, m_g^2 \,
  \Big \langle \big( e \,-\, \eta \big)^2 \big( e \,+\, \eta \big)^2
  \Big\rangle
\Big\}~, 
\label{acmassgrav} 
}
where we have denoted by $g_{\mu \nu}$ the metric associated to the  
vielbein $e_{\mu \nu} \,$. The theory describes a massive graviton and
is characterized by the Planck mass $M_p$ and the graviton mass
$m_g\,$. To write the cosmological constant--like interaction term we
have introduced the notation 
\equ{
\langle A B C D \rangle ~\equiv~
\,-\, \frac{1}{4} \, \epsilon^{\alpha \beta \gamma \delta} \, \epsilon^{a b c d}\, 
A_{\alpha a} B_{\beta b} C_{\gamma c} D_{\delta d}~,
\label{newproduct}
}
with $\epsilon^{abcd}$ being the totally anti--symmetric Levi--Civita
tensor. (Some useful properties of this product are collected in
appendix~\ref{sc:bracket}.) This coupling term is a generalization of
the measure, since   $\langle A^4 \rangle = |A| = \det(A) \,$. Using
these properties it is not hard to show that the action can be
rewritten as 
\equ{
S ~=~ \frac 12 \, M_P^2 \int d^4 x \Big\{
\sqrt{-g} \, \Big( R \, + \, 3 m_g^2 \Big) 
\,-\, \frac 12 \, m_g^2 \,   \big( [e]^2 \,-\, [e^2]  \big)
\Big\}~. 
}
where $[ e^n ]$ is the trace of $( \eta^{-1} \, e )^n \,$. Hence, 
the final term can be interpreted as a FP mass term for the vielbein. 
By formally inverting the definition 
\equ{
{g}_{\mu \nu} ~=~
{e}_{\mu m} \eta^{mn} {e}_{\nu n}~, 
}
we could have equivalently expressed it in terms of the
metric. However, the resulting expression would appear as a
complicated and unmotivated power series of the metric. This model is
obtained by considering a specific limit of the bi--gravity theory
introduced in \cite{gnp} as is explained in appendix~\ref{sc:bigrav}. 
The specific choice of the action (\ref{acmassgrav}) is obtained by
imposing the reflection symmetry  $e_{\gm m} \ra - e_{\gm m}\,$, and
by requiring that the Minkowski background $e_{\mu \nu} = \eta_{\mu\nu}$
 is a solution. Indeed, from this last requirement, we see that the
second term must factorize the $( e - \eta )^2$ combination, and then
the second factor simply follows from the reflection symmetry. This
specifies the action (\ref{acmassgrav}) uniquely among the several
interaction terms which may constructed starting from the vielbein,
the Minkowski background, and the product (\ref{newproduct}).

However, there is still a large arbitrariness in this procedure, given
by the choice of the background term. In the action (\ref{acmassgrav})
we chose it to be equal to the Minkowski background $\eta_{\mu \nu}$
in Cartesian coordinates. But, already choosing it to be the Minkowski
metric in spherical coordinates would result in a different theory,
since this term explicitly breaks covariance. We can also obtain more
general solutions, starting from a different background metric
$e_{b\,\mu \nu}$ in the action (\ref{acmassgrav}), and adding the
corresponding source term which would lead to that solution in the
standard case  since the ``mass term'' is quadratic in $e_{\mu \nu} -
e_{b\,\mu \nu}$, it would not affect the validity of the solution. 
However, for definiteness, in the remainder of this paper we will only
consider the case where the background is Minkowskian using 
Cartesian coordinates.

The interaction term in \eqref{acmassgrav} constitutes a particular
completion of the FP mass term. Inserting the expression 
\equ{
e_{\mu \nu} = \eta_{\mu \nu} + f_{\mu \nu}
}
in the second term of (\ref{acmassgrav}) gives
\equa{
\gD S ~=~  \dsp 
\frac 12 \, m_g^2 M_P^2 \Big\{ \, & 
\Big( [f]^2 \,-\, [f^2]  \Big) 
\,+\, \frac 12 \Big( 
[f]^3 \,-\, 3\, [f] [f^2] \,+\, 2\, [f^3] 
\Big) 
\non \\[2ex] & \dsp 
\,+\, \frac 18 \Big( 
[f]^4 \,-\, 6\, [f]^2 [f^2] \,+\, 3\, [f^2]^2 \,+\, 8\, [f] [f^3] \,-\, 6\, [f^4]
\Big) 
\, \Big\}~. 
\label{acexact}
}
We stress that this expression is exact (rather than just a
perturbative expansion to fourth order). In the remainder of the paper 
we discuss the effects of the nonlinear
interactions in setting the scale at which the model becomes strong.

\section{St\"uckelberg fields}
\label{sc:stuckelberg}

We investigate the graviton interactions in the model outlined
in the previous Section. Nonlinear interactions of massive gravity can
be most easily studied through the St\"uckelberg formalism, developed
in ref.~\cite{ags}. To do so for our model, we first have to formulate
the St\"uckelberg formalism in terms of the vielbein, rather than the
metric (as done in~\cite{ags}).  In this Section we describe this
computation in some detail.

The St\"uckelberg formalism consists in performing a series of
transformations in the action~(\ref{acmassgrav}), and in promoting the
parameters of these transformations to new fields. These fields appear
in the new action together with additional symmetries, so that the
original action can be recovered with a particular gauge choice. 
However, we can also choose alternative gauges in the new action,
where the nonlinear interactions can be computed more easily.

More specifically, we start from a symmetric vielbein  perturbation
$f_{\gm m}$, and perform a general coordinate transformation 
$x \rightarrow y = y(x)$ combined with a local Lorentz transformation,
with $\bL \eta^{-1} \bL^T = \eta \,$. Performing these transformations
into the action (\ref{acmassgrav}) results in the replacements
\equ{
d^4 x \rightarrow d^4 y ~=~ \Big| \pp[y]{x} \Big| \, d^4 x
~, \qquad
 e_{\mu a} \rightarrow e_{\mu a}'
~=~ \frac{\partial x^\alpha}{\partial y^\mu} e_{\alpha b}\, 
 \eta^{bc} \, \bL_{c a}~. 
\label{eprime}
}
We note that, even though $e_{\gm m}$ is assumed to be symmetric,
$e'_{\gm m}$ is certainly not automatically symmetric. While the first
term of \eqref{acmassgrav} is invariant under these combined
transformations, the whole action transforms into the St\"uckelberg form
\equ{
S \rightarrow S_{st} ~=~ \frac{1}{2}\, M_P^2 \int d^4 x
\Big\{
\sqrt{-g} \, R \,+\, 3 \,m_g^2 \,\Big|\pp[y]{x}\Big| \, 
\Big\langle \big( e' \,-\,\eta \big)^2 \big( e' \,+\, \eta \big)^2 \Big\rangle
\Big\}~,
\label{acstuckel}
}
where $e'_{\gm m}$ has been defined in \eqref{eprime}. Notice that 
\eqref{eprime} introduces the Jacobian in front of the cosmological 
constant term.

The parameters in the local Lorentz transformations are not dynamical,
therefore they can be integrated out using their algebraic equations
of motion. Writing 
\begin{equation}
\bL_{mn}(x) ~=~ L_{mp}(x) \Big[ \exp(\get\inv b(x))\Big]^p{}_n~, 
\end{equation}
where $b_{\mu \nu}$ is anti-symmetric, we expand the action
\eqref{acstuckel} to first order in $b\,$:
\equ{
\delta_b S_{st} ~=~ 6 \, m_g^2 M_P^2 \int d^4 x
\, \Big|\pp[y]{x}\Big| \, 
\Big\langle
{e'}^3\,  \big( e' \eta^{-1} b \big)
\,-\, \eta^2 e' \, \big( e' \eta^{-1} b \big)
\Big\rangle~.
\label{variation}
}
Using the property \eqref{newproduct-id1}, it is immediate to verify
that the first term in \eqref{variation} vanishes, due to the
antisymmetry of $b_{mn}\,$.  Using \eqref{newproduct-id2}, the
remaining term rewrites
\equ{
\delta_b S_{st} ~=~ \,-\, \frac 12 \, m_g^2 M_P^2
\int d^4 x \, \Big|\pp[y]{x}\Big| \, 
\Big\{
\big[ e' \big] \big( \eta^{-1} e' \eta^{-1} \big)^{\mu\nu}
\,-\, \big( \eta^{-1} e' \eta^{-1} e' \eta^{-1} \big)^{\mu \nu}
\Big\}
b_{\mu \nu}~,
}
where we have used the notation $[e']$ for the trace, see below
\eqref{acexact}. This contribution vanishes if $e'$ is symmetric, and
therefore $\bL=L\,$ is an on-shell solution.  This symmetry of $e'$
together with the requirement that $L$ is a local Lorentz transformation,
\equ{
 \Big( \frac{\partial x}{\partial y} \,e\, \eta^{-1} L
\Big)^T ~=~ \frac{\partial x}{\partial y} \,e\, \eta^{-1} L
~, \quad {\rm and} \quad
L \eta^{-1} L^T ~=~ \eta~, 
\label{condition}
}
determines $L$ uniquely in terms of the other fields. See
Appendix~\ref{sc:pertexp} for a perturbative construction of $L$. The
freedom in the choice  of its sign is fixed by requiring, that $L=\Id$
if $(\partial x/\partial y) e$ is symmetric.

Therefore, after we solve eqs.~(\ref{condition}) for $L$, and we
substitute the solution back into (\ref{acstuckel}), we are left
with an action which explicitly depends only on the three dynamical
fields $f_{\mu\nu} ,\, a_\mu ,\,$ and $\phi \,$. The latter two fields
are obtained by decomposing  
\equ{
\arry{c}{
y^\gm(x) ~=~ x^\gm \,+\, a^\gm(x) \,+\, \der^\gm \gf(x) 
\;\;,\;\; 
}
\label{perturbations}
}
into the spin--zero, $\gf\,$, and spin--one, $a_\gm\,$, polarizations of
the massive graviton (this decomposition introduces an additional
$U(1)$ gauge symmetry in the theory).  These fields are the starting
point for the St\"uckelberg analysis of the nonlinear interactions
performed in the next Sections.

To summarize the St\"uckelberg formalism we have employed in this
Section: In general, there are more degrees of freedom in the vielbein
than in the metric, which are compensated by the local Lorentz
transformations. This means that in the St\"uckelberg description
additional fields are introduced for both the general coordinate
transformations and the local Lorentz transformation. But contrary to
the St\"uckelberg fields associated with the general coordinate
transformations, the ones of the local Lorentz transformations are
always auxiliary (i.e.\ non--dynamical) fields. When we enforce their
field equations, we ensure that the {\em composite} vielbein
$e'\,$, defined in \eqref{eprime}, remains symmetric {\bf also} after the
St\"uckelberg fields $a_\gm$ and $\gf$ have been introduced. Hence the
number of physical degrees of freedom in the vielbein formulation is
the same as in the metric formulation discussed in~\cite{ags}.

\section{Graviton and St\"uckelberg propagators}
\label{sc:propagators}

The aim of this Section is to compute the propagators of the massive
gravity theory defined in Section \ref{sc:model} using the
St\"uckelberg field decomposition discussed in the previous Section.
As usual the propagators can be read off from the quadratic action
in the perturbations. However, we will encounter a few complications:
First of all, the scalar $\gf$ obtains a regular kinetic term only
after a Weyl rescaling of the graviton field~\cite{ags}. A second
difficulty is that, before inverting the kinetic operator,  we need to
fix the  general coordinate covariance and the additional $U(1)$ gauge
symmetry generated by the St\"uckelberg procedure. Even after the Weyl
rescaling, the tensor $f_{\gm\gn}$, the vector $a_\gm$ and the
scalar $\gf$ mix with each other at the quadratic level (so that they
cannot be used as such to describe independently propagating degrees
of freedom). In the following we show how the last two problems can be
solved together by choosing 't Hooft--Feynman--like gauge fixing
terms.

At quadratic order in $f_{\gm\gn}\,$, $a_\gm \,$, $\gf$ the action
\eqref{acstuckel} is computed using \eqref{newproduct-id2} and the
second order expansion of the matrix $L\inv$ given in \eqref{Linv}.
The result takes the rather complicated form
\equa{
 S_2 ~=&\, \dsp 
 \frac{1}{2}\, M_P^2 \int d^4 x \bigg\{
f^{\mu \nu} \Big[ f_{,\mu \nu} \,+\, \Box f_{\mu \nu}
\,-\, {f^\alpha}_{\mu , \alpha \nu}\,-\, {f^\alpha}_{\nu,\alpha \mu}
\,+\, \eta_{\mu \nu}
\Big( {f_{\alpha\beta,}}^{\alpha\beta} -\Box f \Big)
\Big] 
\label{acquadr}
\\[2ex] \dsp 
& \,+\, m_g^2\, \Big[
 f^2 \,-\, f^{\mu \nu} f_{\mu \nu} 
\,+\,
f^{\mu \nu} \Big( \partial_\mu a_\nu \,+\, \partial_\nu a_\mu \Big)
\,-\, 2 \, f \partial^\mu a_\mu
\,-\, \frac{1}{4} F_{\mu \nu} F^{\mu \nu}
\,+\, 2 \, f^{\mu \nu} \partial_\mu \partial_\nu \phi
\,-\, 2 \, f \Box \phi
\Big]
\bigg\}.
\non 
}
The sign of the kinetic term of the vector $a_\gm$ is the standard
one (had the sign of the mass term in~(\ref{acmassgrav})
been opposite, $a_\gm$ would have been  a ghost). The last two terms
on the second line of \eqref{acquadr} are the only two in which the
scalar $\gf$ appears, and are not regular kinetic terms for a scalar. 
Following~\cite{ags} we perform the linearized Weyl rescaling 
\equ{
f_{\mu \nu} ~=~
{\hat f}_{\mu \nu} \,-\, \frac 12\, m_g^2 \, \eta_{\mu\nu} \,  \phi
\label{Weyl} 
}
of the graviton, to obtain a regular kinetic term for the scalar
$\gf\,$. The quadratic action \eqref{acquadr} becomes
\equ{
S_2 ~=~  \frac 12 \, M_P^2 \!\! \int \!\! d^4 x \bigg\{
{\hat f}^{\mu \nu} \Big[
{\hat f}_{,\mu \nu} \,+\, \Box {\hat f}_{\mu \nu}
\,-\, {\hat f}^\alpha{}_{\mu , \alpha \nu}
\,-\, {\hat f}^\alpha{}_{\nu , \alpha \mu}
\,+\, \eta_{\mu \nu} \Big(
{\hat f}_{\alpha\beta,}{}^{\alpha\beta} - \Box {\hat f}
\Big) \Big]
\,+\, m_g^2 \, \Big[
{\hat f}^2 \,-\, {\hat f}^{\mu \nu} {\hat f}_{\mu \nu}
\Big]
\nonumber \\[2ex]
\,+\, m_g^2 \, \Big[
{\hat f}^{\mu \nu}
\Big( \partial_\mu a_\nu \,+\, \partial_\nu a_\mu \Big)
\,-\, 2 \, {\hat f} \partial^\mu a_\mu
\,-\,\frac{1}{4} \, F_{\mu \nu} F^{\mu \nu}
\,+\, 3 \, m_g^2\,  \phi\Big( \big[\frac 12 \, \Box \,+\, m_g^2\big] \phi
\,+\, \partial^\mu a_\mu \,-\, {\hat f} \Big)
\Big]
\bigg\}~.
\label{achat}
}
Even though we now have obtained a regular kinetic term for the
scalar $\gf$, this action is still rather complicated and contains
quadratic interactions between the three fields $f_{\gm\gn}\,$,
$a_\gm\,$, and $\gf\,$.

In addition, this action \eqref{achat} is invariant under the (linearized)
general coordinate transformations and a $U(1)$ gauge symmetry
\equ{
\gd {\hat f}_{\mu\nu} ~=~
\frac12 \, \big(\varepsilon_{\mu,\nu} \,+\, \varepsilon_{\nu,\mu}\big)
\,-\, \frac12 \, m_g^2\, \eta_{\mu\nu} \, \psi
~,\qquad
\gd a_\mu ~=~ \varepsilon_\mu \,+\, \partial_\mu \psi
~,\qquad
\gd \phi ~=~ \,-\, \psi
~. 
\label{gaugetrans}
}
These gauge symmetries can be fixed by using the following
gauge fixing functionals:
\equ{
{\text{GCC}}~:\quad
\gTh^\gm ~=~
{\hat f}^{\mu\nu},_\nu \,-\, \frac12 {\hat f},^\mu
\,-\, \frac12 \, m_g^2 \, a^\mu
~,\qquad
{U(1)}~:\quad
\gTh ~=~ a_\mu,^\mu \,-\, {\hat f} \,+\, 3 \, m_g^2\, \phi
~.
\label{gaugefunctionals}
}
Indeed, under the combined gauge transformations \eqref{gaugetrans}
these functionals transform as
\equ{
\gd \gTh^\gm ~=~ \frac 12 \, \big( \Box \,-\, m_g^2 \big) \, \gve^\gm
~,\qquad
\gd \gTh ~=~ \big( \Box \,-\, m_g^2 \big) \, \gps
~.
}
Note that at this order $\gTh^\gm$ only transforms under general
coordinate transformations, while $\gTh$ only under the $U(1)$
symmetry. This shows that by suitable gauge transformations these
gauge fixing functionals can be made equal to any prescribed functions. 
When one performs an analysis at the level of equations of motions,
it is most convenient to simply set both functionals $\gTh^\gm$ and
$\gTh$ to zero. However, since in this Section the aim is to obtain
simple forms for the propagators, we employ the gauge fixing
functionals \eqref{gaugefunctionals} to define the gauge fixing action
\equ{
S_{gf} ~=~\,-\, M_P^2 \int \d^4 x \, 
\Big\{ \, \gTh^\gm \gTh_{\gm} \,+\, \frac 14 \, \gTh^2 \Big\}
~.
\label{gaugefix}
}
The use of these gauge fixing functionals can be viewed as a
generalization of the 't Hooft $R_\gx$ gauges in spontaneously
broken gauge theories. In principle, we can allow for arbitrary
normalizations in front of each of the terms in
\eqref{gaugefunctionals}: As long as they are suitably chosen the  
tensor, vector and scalar modes remain decoupled at the quadratic
level. However, in this "Feynman" gauge, the gauge fixed action 
reduces to the simplest form 
\equ{
S_2' ~=~  \frac{M_P^2}2 \!\! \int \!\! d^4 x \Big\{
{\hat f}^{\mu \nu}(\Box -m_g^2)
\Big({\hat f}_{\mu \nu} - \frac12\, \eta_{\mu\nu}{\hat f}\Big)
\,+\, \frac{1}{2}\, m_g^2\, a_\mu (\Box - m_g^2) a^\mu
\,+\, \frac32 \, m_g^4 \, \phi (\Box - m_g^2) \phi \Big]
\Big\}
~.
\label{acdiagonal}
}
Hence in the gauge defined by \eqref{gaugefix} the different spin
states $f_{\gm\gn}\,$, $a_\gm$ and $\gf$ are decoupled, and all have
the same mass. In this gauge all components of these fields are
dynamical.

This action is very convenient: It is immediate from
\eqref{acdiagonal} that the field redefinitions
\equ{
 \hat f_{\gm\gn} ~=~ \frac{1}{M_p}\, \hat f_{c\, \gm\gn}
~,\qquad
 a_\gm ~=~ \sqrt 2\, \frac{1}{m_g M_p}\, a_{c\, \gm}
~,\qquad
 \phi ~=~ \sqrt{\frac23}\, \frac{1}{m_g^2 M_p}\, \phi_c
~,
\label{canonical}
}
normalizes the fields canonically. Also the propagators for the graviton
$\gD_{\gm\gn\,\gr\gs}$, the vector field $\gD_{\gm\,\gn}$ and the
scalar $\gD$ become very simple
\equ{
\Delta_{\ga\gb\, \gr\gs} ~=~
\frac12\, 
\big(
 \eta_{\alpha\rho}\eta_{\beta\sigma} \,+\, \eta_{\alpha\sigma}\eta_{\beta\rho}
\,-\, \eta_{\alpha\beta}\eta_{\rho\sigma}
\big) \, 
\gD
~,\quad
\gD_{\gm\, \gn} ~=~ \get_{\gm\gn} \, \gD
~,\quad
\gD ~=~ \frac i{p^2+m_g^2}
~.
\label{propagators} 
}
These propagators can be obtained from massless propagators in the 
Feynman gauge by the straightforward substitution: $p^2 \ra p^2 +
m_g^2\,$. Even though the determination of the propagators
here was performed in the vielbein formulation, it is obvious that it
extends immediately to metric perturbations $h_{\mu \nu}$ as well,
since at the linearized  level one simply has 
$h_{\mu \nu} = 2 f_{\mu \nu}\,$. Thus, this computation provides
the gauge fixing, which was not explicitly given in ref.~\cite{ags}.

\section{Dominant interactions at high energies}
\label{sc:interactions}

\newcommand{\strongtable}{
\begin{table}
\tabu{cc}{
\(
\arry{|c|c|}{
\hline
\text{4 point} & \text{scale}
\\[1ex] \hline\hline
(\der^2 \gf)^4 & \gL_4 \\[1ex]
(\der^2 \gf)^3 (\der a) & \gL_{3\frac 12} \\[1ex]
(\der^2 \gf)^2 (\der a)^2 & \gL_{3} \\[1ex]
(\der^2 \gf)^3 f & \gL_{3} \\[1ex]
\hline
}
\)
\quad & \quad
\(
\arry{|c c|c|}{
\hline
\text{3 point} & \text{3 point} & \text{scale}
\\[1ex] \hline\hline
(\der^2 \gf)^3 & (\der^2 \gf)^3 & \gL_5 \\[1ex]
(\der^2 \gf)^3 & (\der^2 \gf)^2 (\der a) & \gL_{4\frac 12} \\[1ex]
(\der^2 \gf)^3 & (\der^2 \gf) (\der a)^2 & \gL_{4} \\[1ex]
(\der^2 \gf)^3 & (\der^2 \gf)^2 f & \gL_{4} \\[1ex]
(\der^2 \gf)^3 & (\der a)^3 & \gL_{3\frac 12} \\[1ex]
(\der^2 \gf)^3 & (\der^2 \gf) (\der a) f & \gL_{3\frac 12} \\[1ex]
(\der^2 \gf)^3 & (\der a)^2f & \gL_{3} \\[1ex]
(\der^2 \gf)^3 & (\der^2 \gf) f^2 & \gL_{3} \\[1ex]
\hline
}
\)
~~
\(
\arry{|c c|c|}{
\hline
\text{3 point} & \text{3 point} & \text{scale}
\\[1ex] \hline\hline
(\der^2 \gf)^2 (\der a) & (\der^2 \gf)^2 (\der a) & \gL_4 \\[1ex]
(\der^2 \gf)^2 (\der a) & (\der^2 \gf) (\der a)^2 & \gL_{3\frac12} \\[1ex]
(\der^2 \gf)^2 (\der a) & (\der^2 \gf)^2 f & \gL_{3\frac12} \\[1ex]
(\der^2 \gf)^2 (\der a) &  (\der a)^3 & \gL_3 \\[1ex]
(\der^2 \gf)^2 (\der a) & (\der^2 \gf) (\der a) f & \gL_3 \\[1ex]
(\der^2 \gf)^2 f & (\der^2 \gf)^2 f & \gL_3\\[1ex]
(\der^2 \gf)^2 f & (\der^2 \gf) (\der a)^2 & \gL_3\\[1ex]
(\der^2 \gf) (\der a)^2 & (\der^2 \gf) (\der a)^2 & \gL_3\\[1ex]
\hline
}
\)
\\[2ex]\\
\ref{tb:strong}.a & \ref{tb:strong}.b
\\[3ex]
}
\captn{
This table considers all possible interactions that give rise to
four--point scatterings that becomes strong at scale $\gL_3$ or
below, using canonically rescaled St\"uckelberg fields
\eqref{canonical}. Such scatterings either arise directly from single
four--point functions, \ref{tb:strong}.a,  or by combining two
three--point functions,  \ref{tb:strong}.b, with a scalar,  
vector or graviton propagator in between the two vertices. 
\label{tb:strong}}
\end{table}
}

\strongtable

After the discussion of the quadratic action and propagators, we 
now turn to the interactions of this massive gravity theory. The
interactions of any massive gravity theory are rather involved,  
because there are many of them, and they all possess a rather
complicated tensorial structures. Moreover, the interactions have a
polynomial momentum dependence, and become non--perturbative 
at some ``large'' energy scale (in this context, ``large'' is defined
with respect to the graviton mass). Therefore it is useful to classify
which interactions are the dominant ones at high energies.

As we reviewed in Section \ref{sc:stuckelberg}, in the St\"uckelberg
formalism the vector $a_\gm$ enters in the ``pion'' field with one
derivative, while the scalar $\phi$ with two derivatives. Therefore,
barring cancellations, the scalar couplings will be in general the
largest at high energies. As was noted in~\cite{ags}, the
$\gf\gf\ra\gf\gf$ scattering generally becomes strong at the scale  
$\Lambda_5= (m_g^4 M_P)^{1/5}\,$. This scale falls into the regular  
pattern of scales defined by 
\equ{
\gL_p ~=~ \Big( m_g^{p-1} M_P \Big)^{1/p}
~,
\label{gLp}
}
where $p$ can be integer or half integer (notice that $\Lambda_p$ is 
a decreasing function of $p$). All interactions that become strong at
scale $\gL_3$ or below are grouped in table \ref{tb:strong}. In
particular, the scale $\gL_5$ appears in $\gf\gf\ra\gf\gf$ scattering
by combining two three--scalar interactions. It was also shown
in~\cite{ags} that the scale at which this scattering becomes strong
can at most be raised to  $\Lambda_3\,$. For a generic model of
massive gravity, they sketched a procedure to obtain a suitable set 
of counter terms.  We show that the model we are considering
automatically satisfies this property.

To give a detailed discussion, we divide the presentation in three
Subsections. The first one proves that our model does not have any
interactions with either only scalars or at most one vector. In the
second Subsection we show that all tree level processes become 
strong at a scale which is greater or equal to $\gL_3$. In the final
Subsection we explicitly compute $\gf\gf \ra\gf\gf$ scattering, as an
example, and show that it is possible to extend the model such that
this amplitude vanishes altogether.

\subsection{Absence of interactions with only scalars or at most one vector}
\label{sc:34point} 

In the next two Subsections we want to show that our theory does not
become strong below the scale $\gL_3$. Here we prove that the most
dangerous interactions, which have the schematic forms
$(\partial^2\phi)^n$ and $(\der^2 \gf)^n (\der a)\,$, are all absent
in our model. In these subsections we consider the scalar interactions
before Weyl rescaling: The additional scalar interactions introduced
by the Weyl rescaling will have an additional factor $m_g^2$ from
\eqref{Weyl}, and have therefore the same strength as tensor
interactions. The interaction term of \eqref{acstuckel} can be written
as   
\equ{
\gD S_{st} ~=~
\frac 32\,  m_g^2 M_P^2 \int d^4 x \, 
\Big\langle
\big( e \,-\, \Pi \big)^2 \big( e \,+\, \Pi \big)^2 \rangle
~,
\label{accosmo}
}
where we have introduced the short-hand notation
\equ{ \Pi_{\mu a} ~=~
\frac{\partial y^\nu}{\partial x^\mu} \, 
\eta_{\nu b} \left( L^{-1} \right)^{b c} \, \eta_{c a}
~,
\label{matrixPi}
}
where $y^\nu$ is expressed by \eqref{perturbations} in terms of
$a_\gm$ and $\gf$. In the following, it is often useful to be able to
resort to matrix notation to suppress the indices. Aside from the
Minkowski metric $\get$, its perturbation $f$ and the Lorentz
transformation $L$ (which are matrices by definition), we encode the
derivatives of $y$ into the matrices 
\equ{
\gp ~=~ \Big( \pp[y]{x} \,-\, \Id \Big) \, \get ~=~
\gF \,+\, \frac 12 \,\Big( A \,+\, A^T \Big) \,-\, \frac 12\, F
~,
\label{matrixpi}
}
where $\gF_{\gm\gn} = \gf_{,\gm\gn}\,$, $A_{\gm\gn} = a_{\gm,\gn}$ 
and $F_{\gm\gn} = \partial_\gm a_\gn - \partial_\gn a_\gm$ is the
$U(1)$ field strength. (For us $\pi$ is defined strictly by the matrix
equation \eqref{matrixpi}; our definition differs slightly from the 
conventions used in ref.~\cite{ags}.)

Now let us first consider interactions with only scalar fields. 
To single out from \eqref{accosmo} the interactions $(\der^2\gf)^n$,
we can simply replace $\gP_{\gm\gn} \ra \get_{\gm\gn} + \gF_{\gm\gn}$
and set both $L_{\gm\gn}$ and $e_{\gm\gn}$ equal to the Minkowski metric
$\get_{\gm\gn}\,$ (ignoring the vector and tensor contributions). 
We can take $L_{\gm\gn} = \get_{\gm\gn}$ because in this case $e'$ is
automatically symmetric. This gives 
\equ{
\gD S_{st} \supset 3 \, m_g^2 M_P^2 \int d^4 x
\big\langle
4 \, \get^2 \,\gF^2 \,+\, 4\,  \get\, \gF^3 \,+\, \gF^4
\big\rangle~, 
\label{acsimple}
}
where we have employed the matrix notation defined in
\eqref{matrixpi}. All these terms vanish upon partial integration: For
example, for the first term we obtain
\equ{
\int d^4 x \, \big\langle 4 \,\get^2 \,\gF^2 \big\rangle ~=~
\,-\, \int d^4 x \, \ge^{abcd} \ge_{ab}{}^{\gr\gs} \,\gf_{,c\gr} \gf_{,d\gs} ~=~
 \int d^4 x \,\ge^{abcd} \ge_{ab}{}^{\gr\gs} \,\gf_{,\gr} \gf_{,cd\gs}
~=~ 0 ~,
}
where we have used the definition \eqref{newproduct} and the
anti--symmetry of the $\ge^{abcd}$ tensor. Similar arguments also
apply to the other two terms. Hence, all the interactions of the form
$(\der^2 \gf)^n\,$ vanish. Equivalently this result can be
obtained by going to momentum space and realizing that then all
matrices are of the form \eqref{vectormat} for which the angular
bracket vanishes as is proven in Appendix \ref{sc:bracket}. As
anticipated below \eqref{acexact}, this shows that  all these pure scalar
higher derivative interactions naturally vanish, which was the 
motivation in ref.~\cite{crem} to consider these combinations.

Also the interactions of the schematic form $(\der^2 \gf)^n (\der a)$
vanish. To see this, we employ the matrix notation 
\eqref{matrixpi}, and we again set $e=\get$ in the interaction term
\eqref{accosmo}. We first observe that all the terms which are linear
in $F$ vanish. This is due to the fact, that  all the other tensorial
structures which would multiply $F$, namely $\get$ or $\gF$, are
symmetric, while $F$ itself is anti--symmetric. Therefore, we can set
both $f_{\mu \nu}$ and $F_{\mu \nu}$ to zero, without losing the term
that we are looking for. Doing so, we have a symmetric pion matrix,  
$\gp = \gF + \frac 12 A + \frac 12 A^T\,$. Once we insert it in
eq.~(\ref{matrixpi}), we obtain a symmetric $e'$ vielbein already,
therefore $L = \eta$. The expression we are looking for can be found
by evaluating eq.~(\ref{accosmo}) for 
$F_{\mu \nu} = f_{\mu \nu} = 0 \;,\; L_{\mu \nu} = \eta_{\mu \nu} \,$.
This results in an interaction action which is again of the form
(\ref{acsimple}), upon the replacement 
$\gF \ra \gF + \frac 12 A + \frac 12 A^T$. The final step is to single
out from this expression the terms which are linear in the vector
field. It is clear that they are of the form $\langle A \gF^n
\get^{3-n}\rangle\,$. Using a similar partial integration procedure as
for the terms $\langle\gF^n\get^{4-n}\rangle$ presented above (or
again using \eqref{vectormat} in momentum space), it is easy to verify
that also such terms vanish.

\subsection{All interactions become strong at scale $\boldsymbol{\gL_3}$ or
above}
\label{sc:npoint}

We now show that there are no scattering amplitudes that become strong
at a smaller scale than $\gL_3$. We only analyze the theory at
the classical level, in particular, we do not consider loop graphs.
We first consider the $S$--matrix elements that correspond
to diagrams that contain a single vertex $V_{n_g,n_a,n_\gf}$, with
$n_g$ tensor, $n_a$ vector and $n_\gf$ scalar external
legs.~\footnote{We would like to remind the reader that, for reasons 
explained in the previous subsection, we count the scalar and tensor
legs before Weyl rescaling.} Taking into account that at a vertex
there is momentum conservation encoded in a single overall four
dimensional momentum delta function, which scales as $1/E^4$ (where
$E$ is the energy in the scattering), we see that such vertex scales
as   
\equ{
V_{n_g, n_a, n_\gf} ~\sim~ 
\frac {m_g^2 M_P^2}{E^4} \, 
\Big( \frac{1}{M_P} \Big)^{n_g} \, 
\Big( \frac{E}{m_g M_P} \Big)^{n_a} \, 
\Big( \frac{E^2}{m_g^2 M_P} \Big)^{n_\gf}~. 
}
where the $E$ in parenthesis arise from derivatives (scalars enters in
the St\"uckelberg formalism with two derivatives, while vectors with
one), and the denominators arise from the canonical normalization. To
understand at which energy $E$ such process becomes strong we need to
compute the scaling of the corresponding $S$--matrix element  
$S_{n_g,n_a,n_\gf}$, in which we integrate over all possible external 
momenta. Because the external particles are all on--shell, we
integrate over  
\equ{
\int d^4 p_i \gd(p_i^2 - m_i^2) ~=~ 
\int \frac { d^3 \vec p_i }{2 E_{i\, \vec p_i}}
~\sim~ 
E^2~, 
}
where in the last step we used that in the high energy regime we are 
considering $E >> m_g \,$. Hence the corresponding $S$--matrix 
element scales like  
\equ{
S^{1/2}_{n_g, n_a, n_\gf} ~\sim~ 
\Big( \frac {m_g M_P}{E^2} \Big)^2 \, 
\Big( \frac{E}{M_P} \Big)^{n_g} \, 
\Big( \frac{E^2}{m_g M_P} \Big)^{n_a} \, 
\Big( \frac{E^3}{m_g^2 M_P} \Big)^{n_\gf}~. 
\label{gens}
}
(For notational convenience we have taken the square root of the
$S$--matrix element; the interaction becomes strong when the
matrix element, or its square root, exceed one). 

Now we analyze the strength of general interactions with a single
vertex. Because we are interested in scatterings, the total number of
external legs $n_g+n_a+n_\gf \geq 3\,$. In total we can distinguish
five different cases: 
(i) $n_a=n_g=0\,$,
(ii) $n_a=n_\gf=0\,$, 
(iii) $n_g, n_\gf \geq 1\,$, 
(iv) $n_a=1\,$  
and finally 
(v) $n_a \geq 2\,$. 
%
%
We have shown in the previous Subsection, that the model does not have 
vertices with only scalars, hence there are no processes that
correspond to the first case (i).  
%
%
The second case (ii) involves interactions among only tensors. The
corresponding $S$--matrix elements
\equ{
S_{n_g , n_a = n_\phi = 0}^{1/2} ~\sim~ 
\Big( \frac {m_g}{M_P} \Big)^2 
\, 
\Big( \frac{E}{M_p} \Big)^{n_g-4}~
}
are bounded from above for $n_g \leq 4\,$, and therefore do not lead
to strong coupling behavior. For $n_g \geq 5$ these interactions 
become strong at an energy scale, which is greater than the Planck 
scale $M_P$.  
%
%
Next, let us consider interactions of type (iii) which
involves at least one scalar and one graviton. By rewriting the
general expression~(\ref{gens}) as 
\equ{
S^{1/2}_{n_g \geq 1, n_a, n_\gf \geq 1} ~\sim~ 
\Big( \frac{E}{\gL_1} \Big)^{n_g-1} \, 
\Big( \frac{E}{\gL_2} \Big)^{2 n_a} \, 
\Big( \frac{E}{\gL_3} \Big)^{3(n_\gf -1)}
~, 
} 
we conclude that these interactions are still weak at energy scales
lower than $\gL_3\,$, because all the exponents of the factors are
nonnegative, and the scales $\gL_3 < \gL_2 < \gL_1= M_P\,$ are ordered
hierarchically, see their definitions \eqref{gLp}. 
%
%
Let us turn to interactions of type (iv), with a single vector,
$n_a=1\,$. From the previous Subsection we know that if there are no
tensors, $n_g=0\,$, the amplitude vanishes. The situation with at
least one tensor and one scalar constitutes a special case of
(iii). The remaining possibility of type (iv), no scalars and $n_g
\geq 2\,$, becomes strong above the Planck scale $M_P$ (analogously to
case (ii)). This is clear when we rewrite
\equ{
S_{n_g \geq 2, n_a =1, n_\phi = 0}^{1/2} ~\sim~ 
\Big( \frac {m_g}{M_P} \Big)^2 
\, 
\Big( \frac{E}{M_p} \Big)^{n_g-2}~. 
}
%
%
Finally, also case (v) remains weak at all scales lower
than $\Lambda_3 \,$, because we can write: 
\equ{
S^{1/2}_{n_g, n_a \geq 2, n_\gf} ~\sim~ 
\Big( \frac{E}{\gL_1} \Big)^{n_g} \, 
\Big( \frac{E}{\gL_2} \Big)^{2(n_a-2)} \, 
\Big( \frac{E}{\gL_3} \Big)^{3 n_\gf}~. 
} 
Hence, we showed that, in this model, all $n-$leg interactions with a
single vertex become strong at the scale $\Lambda_3$ or higher.

Finally, let us discuss tree diagrams that contain more than one
vertex. Such diagrams can be obtained recursively by combining tree
level diagrams with less vertices inside. Whenever we combine two
such diagrams, we loose two external lines, and hence two 
factors of $E^2\,$ in the $S$--matrix. At the same time we gain a
factor $E^2$ in the amplitude, since we have an additional momentum
integral over a propagator: $\int d^4p/p^2 \sim E^2$. (Because we
consider tree diagrams all momenta inside a given diagram are fixed by
momentum conservation.) Therefore, the scaling of the $S$--matrix
element of the combined diagram is the same as of the original two
disconnected diagrams. We can repeat this argument recursively, as we
split any tree level diagram in a series of single vertex
diagrams. Since we already saw that all single vertex scatterings become
strong at least at the scale $\gL_3\,$, this argument shows that this
is the case for any tree level scatterings.~\footnote{All the arguments
in this Section ignore any possible interference between different
diagrams. Interferences can only soften the scattering amplitudes, so
they do not affect the conclusion that the interactions are weak at
energies smaller than $\Lambda_3 \,$.}

\subsection{Can the $\boldsymbol{\gf\gf \ra \gf\gf}$ scattering amplitude vanish?}
\label{sc:phi2raphi2}

\begin{figure}
\begin{center}
\tabu{ccccc}{
   \scalebox{1}{ \epsfig{file=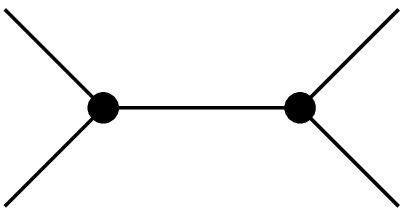} }
& \quad  & 
   \scalebox{1}{ \epsfig{file=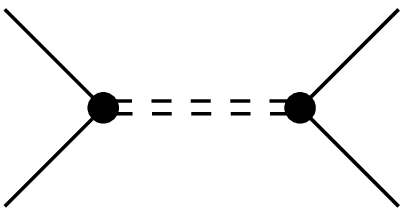} }
& \quad  & 
   \scalebox{1}{  \epsfig{file=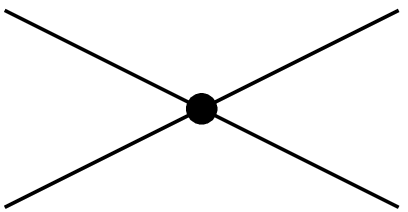} }
\\[2ex]
\ref{fg:gf4scatt}.a & & \ref{fg:gf4scatt}.b &&  \ref{fg:gf4scatt}.c 
}
\end{center}
\captn{
This diagram displays the three diagrams that contribute to the
four scalar scattering at the scale $\gL_3$, which result from
\eqref{intergL3}. In the first diagram the
scalar $\gf$ is exchanged, while in the second diagram the graviton
$\hat f_{\gm\gn}$ is the mediating particle.  The last diagram results
from the four point interactions of scalars. 
\label{fg:gf4scatt}}
\end{figure}

The interaction vertices given in Subsection \ref{sc:34point} can be
employed to obtain several four--point amplitudes which (if not
vanishing) all become strong at the scale $\gL_3$. These amplitudes
correspond to the scattering processes 
$\phi\phi \ra \phi\phi\,$,  $\phi \hat f \ra \phi \hat f\,$, $a\phi
\ra a\phi\,$, $aa \ra aa\,$, and $\gf \hat f \ra a a\,$, plus the
crossed processes. As an example, we compute the 
$\phi\phi \ra \phi\phi$ scattering at tree level. In particular, we want to
investigate whether it is possible to have a model where the full
leading $\phi\phi \ra \phi\phi$ scattering vanishes at tree level.

As will become clear below, the amplitude for $\phi\phi \ra \phi\phi$ does not
vanish in the model \eqref{acmassgrav}. Therefore, we consider the slightly
generalized interaction term  
\equ{
\gD S_{gen} ~=~ \frac {6 m_g^2 M_P^2}{4+\ga+\gb} \int d^4 x 
\Big \langle 
\big( e - \get \big)^2 
\big( 
e^2 \,+\, (2+\ga) \get \,e \,+\, (1+\gb) \get^2 
\big) 
\Big \rangle~,
\label{modac} 
}
where the real parameters $\ga$ and $\gb$ are arbitrary. The
normalization factor $4+\ga+\gb$ in \eqref{modac} is chosen such that
$m_g^2$ still represents the graviton mass. Note that for $\ga=\gb=0$
we recover the terms in eq.~\eqref{acmassgrav}, which is symmetric under
reflection of the vielbein. (Also in this more general model, the
interactions which can potentially become strong at a scale lower than
$\Lambda_3$ vanish for the same reasons we have discussed in the
previous Subsections.) To be able to directly compare various
scattering amplitudes, we use the canonically normalized fields
defined in \eqref{canonical}. The leading interactions of the action
\eqref{modac} can be expressed as 
\equa{&
\gD L_{gen} ~\supset~ 
\,-\, \frac 1{\sqrt{6}} \,\frac{1}{m_g^2 M_p}\, 
\Big\{
 \frac {2+3b}{2} \, [F_c^2\,\gF_c] \,-\, \frac{1+3b}{4} \, [F_c F_c] \, \Box \phi_c 
\Big\}
\non \\[2ex] &
\,+\, \frac 1{3} \, \frac{1+3b}{m_g^2 M_p} \, 
\Big\{
\frac 1{\sqrt 6}\, 
\phi_c\, 
\Big( (\Box\phi_c)^2 \,-\, [\gF_{c}^2] \Big)
\,-\, 
 \frac12\, [\hat f_c] \, (\Box \phi_c)^2 \,+\,[\hat f_c\gF_c^2]
 \,-\,\frac12\, [\hat f_c]\, [\gF_c^2] \,-\,[\hat f_c\gF_c] \, \Box\phi_c
 \Big\}
\non \\[2ex] &
\,+\, \frac {a-2b}{m_g^4 M_P^2} 
\Big\{ 
\frac 1{3\sqrt{6}} 
\Big( \! 
(\Box \gf_c)^3 [\hat f_c] 
\,-\, 3\Box \gf_c\, [\hat f_c]\, [\gF_c^2] 
\,-\, 3 (\Box \gf_c)^2 \, [\gF_c \hat f_c] 
\,+\, 3\, [\gF_c^2]\, [\gF_c\hat f_c] 
\,+\, 2\, [\hat f_c]\,[\gF_c^3] 
\,-\, 6\, [\hat f_c \gF_c^3]
\non \\[2ex] &
\qquad \qquad \quad 
\,+\, 6\, \Box \gf_c\, [\hat f_c \gF_c^2] 
\Big) 
\,+\, \frac{1}8 \Big( [\gF_c^2] [F_c^2] \,-\, (\Box \gf_c)^2 [F_c]^2 \Big)
\,-\, 
\frac 1{18} \, \gf_c 
\Big(
(\Box \gf_c)^3 \,-\, 3 \Box \gf_c\, [\gF_c^2] \,+\, 2\, [\gF_c^3] 
\Big)
\Big\}
\non \\[2ex] & 
\,+\, \frac 14\, 
\frac{1}{m_g^4 M_p^2}\, 
\Big\{
\Big( \frac 23 + 5 b - 2 a \Big) [\gF_c^2 F_c^2] 
\,+\, \Big( \frac 23 + 3 b - a \Big) [\gF_c F_c \gF_c F_c] 
\,-\, \Big(\frac 13  +5 b - 2 a  \Big) \Box \gf_c [\gF_c F_c^2]
\Big\}
\,, 
\label{intergL3}
}
where we have defined the parameters 
\equ{
a ~=~ \frac{\ga}{4 \,+\, \ga \,+\, \gb}~, 
\qquad 
b ~=~ \frac{\gb}{4 \,+\, \ga \,+\, \gb}~.
}
To obtain these interactions we first performed a St\"uckelberg
transformation to the action \eqref{modac}, we then expanded it 
up to first order in $f$ and fourth order in $\gp$,
and we finally performed the linearized Weyl rescaling. To single out
only the leading terms, we then substituted $\gp = A + \gF$ and kept
only those terms with the highest power of $\gF$, since they contain
the greatest number of derivatives. In this way, we found the expansion
\eqref{leadingPi} given in Appendix \ref{sc:pertexp} for $\gP$
defined in  \eqref{matrixPi}. Finally we worked out the brackets
$\langle\ldots\rangle$ in terms of traces $[\ldots]\,$, using the
identities of Appendix \ref{sc:bracket}.

This fourth order leading expression is rather involved for generic 
values of $a$ and $b$. However, notice that if $a = 2b$ the entire
third and fourth lines of \eqref{intergL3} vanish. In particular, 
there are no pure scalar interactions anymore. This is also the case
for the model \eqref{acmassgrav}, with $a=b=0$, which we are mostly
concerned with in this paper. However, if in addition $b = -1/3$, also
the second line and the last term on the last line vanishes, and only
three different interactions survive: one two--scalar vector
interaction and two two--scalar two--vector interactions. In
particular there are no interactions with tensors left. Hence the
only possible four--point scatterings involve two scalars and two
vectors.

We compute the leading tree level $\gf\gf\ra\gf\gf$ scattering
which becomes strong at the scale $\Lambda_3\,$ for generic values of
$a$ and $b$, using the leading expansion \eqref{intergL3}. The process
is described by the three diagrams shown in figure~\ref{fg:gf4scatt}. 
The computation of these diagrams is a straightforward exercise in
Feynman graph computations and therefore only the result is given
here. We neglect the graviton mass against the external and internal
momenta of the scattering. In terms of the standard Mandelstam
variables $s$, $t$ and $u$, the amplitude reads
\begin{equation}
\mathcal{M}(\phi\phi\ra \phi\phi) ~=~
\left[ - \frac{\left( 1 + 3 b \right)^2}{8} + \frac{\left( 1 + 3 b \right)^2}{12} +
\frac{a - 2 b }{3} \right] \frac {stu}{m_g^4 M_P^2}~. 
\end{equation}
The three terms correspond to the diagrams \ref{fg:gf4scatt}.a,
\ref{fg:gf4scatt}.b, and \ref{fg:gf4scatt}.c, given in figure
\ref{fg:gf4scatt}, respectively. The total amplitude expressed in
terms of the center of mass energy $E$ equals 
\equ{
\mathcal{M}(\phi\phi\rightarrow \phi\phi) ~=~ \,-\, \frac 23 
\Big[ 
8 (2\, b\,-\, a) \, + \, 
(1\,+\, 3\,b)^2 
\Big] 
\frac{E^6}{M_p^2 m_g^4} \, \sin^2 \theta
~,
\label{phiphiamplitude}
}
where $\gth$ defines the angle between the momenta of an ingoing and
an outgoing particle in the center of mass frame. 
We see that, besides the special point $a = 2 b = -2/3$,  there is
a whole parabola $a(b) = 2b + (1+3b)^2/8$ such  
that the full tree level $\gf\gf\ra\gf\gf$ scattering vanishes at the scale
$\gL_3\,$. We emphasize that this four--point scalar scattering alone is not
sufficient to understand the strong coupling dynamics of massive
gravity theories, and of our model in particular.

\section{Conclusions}
\label{sc:conclusions}

The model of massive gravity considered in this paper is standard
general relativity with a cosmological constant and a Fierz--Pauli 
mass term for the vielbein. By requiring a reflection
symmetry of the vielbein, $e_{\gm a} \ra -e_{\gm a}$, this theory, like
standard Fierz--Pauli massive gravity, is described by only two parameters:
the Planck scale and the graviton mass. As such it constitutes one of
the simplest non--linear extension of massive gravity theories. This
model can be obtained from a bi--gravity theory with an extremely simple
interaction term between the two sectors. The bi--gravity theory contains
a massless and a massive graviton state, and it admits a
limit in which the massless graviton decouples~\cite{gnp}, as we show in
appendix~\ref{sc:bigrav}.  In the present work we studied the model of a
single massive graviton that emerges in this limit.

We have investigated whether the simplicity of the
model has some physical implications, or only has aesthetic
merit. To do so, we studied this model at the non--linear level
(since, at the linear level, it is equal to the standard Fierz--Pauli
model). As the present model is a special type of massive gravity, 
it shares the problems that massive gravity theories have 
in general. 
The main one is that they are plagued by ghosts at the the non--linear
level~\cite{bd,gagr}, which can never be avoided at the quartic order
in the perturbations~\cite{crem}. A related difficulty is that the
self--interactions of a massive graviton become strong at macroscopic
distances from a source~\cite{arkady1}. In general, the scattering
becomes strong at the energy scale 
$\smash{\Lambda_5 \equiv \left( m_g^4 M_p\right)^{1/5}}\,$. 
However,  by adding to the starting theory a set of suitable nonlinear
interactions.  this scale can be raised up to 
$\smash{\Lambda_3 \equiv \left( m_g^2 M_p \right)^{1/3}}\,$  
at most~\cite{ags}. Even though our theory does not solve these
essential problems of massive gravity theory, it provides a minimal
model that has self--interactions which become strong at the highest
possible scale $\Lambda_3 \,$. Therefore, the model here discussed may
be considered as a prototype of massive gravity, since, in addition to
its minimal formulation, has the best behavior that we can hope to
obtain for such theories.

We conclude with a note on the use of the vielbein rather than the
metric formulation. In eq.~(\ref{acexact}), we wrote the (exact)
lagrangian of the model in terms of metric
perturbations. Ref.~\cite{crem} already showed that special 
combinations of terms have the consequence of removing from the theory
the self--interactions which involve only the scalar polarization of
the graviton, before Weyl rescaling (such interactions - if present -
would lower the strong scale below $\Lambda_3 \,$). However, what in
the metric computation appears only as a computational result
(obtained by allowing for arbitrary coefficients, and then finding
which combination eliminates the unwanted terms), in the vielbein
formulation corresponds to one of the simplest combination of a 
Fierz--Pauli mass for vielbein perturbations and a cosmological constant. 
This suggests that, despite it is very rarely considered, the vielbein 
approach may be more suitable for the study of massive gravity, and, 
hopefully, for finding the improvements that it still requires.

\subsection*{Acknowledgments}

The work of M.P.\ and M.S.\ was partially supported by DOE grant
DE-FG02-94ER-40823, and by a grant from the Office of the Dean
of the Graduate School at the University of Minnesota.

\appendix
\begin{center}
\section*{Appendixes}
\end{center}

\def\theequation{\thesection.\arabic{equation}}
\setcounter{equation}{0}

\section{Derivation from a bi--gravity theory}
\label{sc:bigrav}

In this appendix we explain how the massive gravity model discussed in
this paper can be derived from the bi--gravity theory discussed
in~\cite{gnp}. One starts from the two metrics $\tg_{+\, \gm\gn}$ and
$\tg_{-\, \gm\gn}$, which we rewrite in terms of
two vielbeins $\te_{+\,\gm m}$ and $\te_{-\,\gm m}$, using the
standard definition
\equ{
{\tilde g}_{\pm \,\mu \nu} ~=~
{\tilde e}_{\pm \,\mu m} \eta^{mn} {\tilde e}_{\pm \,\nu n}~.
\label{vielbein}
}
The action for the model is
\equ{
S_{bi} ~=~ \int d^4 x
\Big\{
\sqrt{- \, {\tilde g}_+} \, \Big( \frac{1}{2}M_+^2 \,
  {\tilde R}_+  \,+\,  \Lambda_+ \Big)
+  \sqrt{- \, {\tilde g}_-}\, \Big( \frac{1}{2}M_-^2 \,
   {\tilde R}_- \, + \, \Lambda_- \Big)
\,-\, 2 \, \Lambda_0 \,
 \langle {\tilde e}_+^2 \, {\tilde e}_-^2 \rangle
\Big\} ~. 
\label{acbigravity} 
}
The two gravitational sectors are characterized by the two `'Planck
masses'' $M_\pm$ and the cosmological constants $\gL_0$ and
$\gL_\pm$.~\footnote{As we see below $\gL_0$ needs to be positive to
avoid a tachyonic mass for the graviton.} The double covariance of the
model is broken by the last term.

Let us now proceed to the study of the spectrum of the model for the
Minkowski background. We set
\equ{
e_{\pm\, \mu \nu} ~=~ \eta_{\mu \nu} \,+\, f_{\pm\, \mu \nu}~, 
\label{expapm}
}
and we expand the action~(\ref{acbigravity}) at the quadratic level in
the perturbations $f_{\pm\,\gm\gn}$. The resulting action is not
diagonal in terms of these two modes; however, it can be 
diagonalized through the redefinition 
\equ{
\pmtrx{ f_{+\,\gm\gn} \\ f_{-\,\gm\gn} }
~=~ \frac{1}{r+\frac{1}{r}}
\pmtrx{ 1 & - \frac{1}{r} \\  1 & r }
\pmtrx{ f_{0\,\gm\gn} \\f_{\gm\gn} }
~, \qquad
r ~=~ \frac{M_+}{M_-} c^2 
~. 
\label{diagonalization}
}
The mode $f_{0\,\gm\gn}$ is massless, while $f_{\gm\gn}$ has a
Fierz--Pauli mass $m_g$, which is related to the cosmological constant
$\gL_0$ by \footnote{We correct a typo in the definition of the mass
appearing in~\cite{gnp}.}
\equ{
\gL_0 ~=~ \frac 32 \, m_g^2 \, M_P^2~,
\qquad
M_P^2 ~=~ M_+ M_- \,\Big( r \,+\, \frac 1r \Big)\inv~.
\label{mgmp}
}
The Planck mass $M_P$ (as obtained from the kinetic terms) is found to
be identical for both gravitons. The graviton mass 
$m_g$ vanishes for $\gL_0 = 0\,$, as  a consequence of the enlarged
covariance.

The massless graviton decouples in the limit of either
$M_+ \rightarrow \infty$ (with finite $M_-$), or $M_- \rightarrow
\infty$ (with finite $M_+$). Both the Planck and graviton mass
appearing in~(\ref{mgmp}) are finite in these limits, so that the
quadratic actions for the massless and massive modes remain
finite. However, all the nonlinear interactions involving the massless
mode $f_{0 \mu \nu}$ vanish in either limit. For instance, in the
first limit, we have at leading order $f_+ \propto f_0 / M_+ \,, f$,
and $f_- \propto f_0 / M_+  \,, f / M_+^2 \,$ (suppressing for
shortness the tensorial indices). Therefore, writing the original
action~(\ref{acbigravity}) in terms of $f$ and $f_0$, we see that any
term containing the massless mode is suppressed by a negative 
power of $M_+$, and therefore vanishes as 
$M_+ \rightarrow \infty\,$. The only exception is the quadratic
kinetic term from the $+$ sector, which results in a finite kinetic
term for the massless graviton with the Planck mass~(\ref{mgmp}). In
this limit, the massive graviton is identified with $f_-$, and a
theory of a single massive graviton is obtained by coupling all 
matter fields in the $-$ sector. Obviously, an analogous situation is
obtained in the other limit.

To obtain the model of this paper, we study the theory in either of
the two (equivalent) limits, by ignoring the decoupled massless
graviton. We also restrict our attention to the Minkowski
background. For instance, in the  
$M_+\rightarrow\infty$ limit, this gives 
${\tilde e}_{+ \, \mu \nu} = c \, \eta_{\mu\nu}\,$,  
${\tilde e}_{- \, \mu \nu} = c^{-1} e_{\mu \nu} = c^{-1}
( \eta_{\mu \nu} + f_{\mu \nu}) \,$, and the
action~(\ref{acbigravity}) reduces to~\eqref{acmassgrav}.

\section{Properties of $\boldsymbol{\langle ABCD \rangle}$}
\label{sc:bracket}

In this Appendix we collect various helpful properties of the angular
bracket $\langle ABCD \rangle$ defined in eq.~(\ref{newproduct}). 
First of all the ordering of the matrices $A,\ldots,D$ is irrelevant in this 
expression, because of the two Levi--Civita tensors in its
definition. For the same reason, this product is invariant under the
simultaneous transposition of all four matrices. 
As we remarked in the main text, it generalizes the notion
of a determinant, in the sense that $\langle A^4 \rangle = \det(A)
= |A| \,$. However, $\langle A^2 B^2\rangle$ cannot be written as
a determinant.  The property of a determinant, that the determinant of
a product of matrices is equal to the product of their determinants,
generalizes to  
\equ{
\big\langle \, (aAb) \, (aBb) \, (aCb) \, (aDb) \,\big\rangle ~=~
|a| \, \langle ABCD \rangle \, |b|
~, 
\label{newproduct-id1}
}
for any matrices $a,b,A,B,C,D\,$.

The angular bracket can be rewritten in terms of traces 
$\left[ A \right] = \eta^{ab} A_{ba}\,$; the resulting expression is 
rather involved
\equ{
\langle ABCD \rangle ~=~ \frac 1{24} \, \Big(
[A] \, [B] \, [C] \, [D] 
\,-\, [A] \, [B] \, [CD] \,-\, [A] \, [D] \, [BC] \,-\, [B] \, [D] \, [AC] 
\,-\,[AB] \, [C] \, [D] 
\non \\[2ex] 
\,-\, [A] \, [BD] \, [C] \,-\, [AD] \, [B] \, [C] 
\,+\, [AB] \, [CD] \,+\, [AC] \, [BD] \,+\, [AD] \, [BC] 
\,+\, [A] [BCD]
\non \\[2ex] 
 \,+\, [A] \, [BDC] \,+\, [B] \, [ADC] \,+\, [B] \, [ACD] 
\,+\, [C] \, [ABD] \,+\, [C] \, [ADB] \,+\, [D] \, [ABC] 
\non \\[2ex] 
\,+\, [D] \, [ACB]
\,-\, [ABCD] \,-\, [ADBC] \,-\, [ACDB]
\,-\, [ABDC] \,-\, [ACBD] \,-\, [ADCB]
\Big)~. 
\label{full} 
} 
However, when some of its entries are equal to the Minkowski metric 
$\eta\,$, its expression simplifies considerably:  
\equ{
\langle\, \eta^4 \,\rangle ~=~  1
~,\qquad 
\langle\, \eta^3 \,A \,\rangle ~=~  \frac 14 \, [A]
~,\qquad 
\langle\, \eta^2 \,A B \,\rangle ~=~ 
\frac1{12} \, \Big( [ A] \, [ B ] \,-\, [AB] \Big)~, 
\non \\[2ex] 
\langle\, \eta \, ABC \,\rangle ~=~  \frac 1{24}\, \Big( 
[A] \, [B] \, [C] \,-\, [A] \, [BC] \,-\, [B] \, [AC] \,-\, [C] \, [AB]
 \,+\, [ABC] \,+\, [ACB] 
\Big)~. 
\label{newproduct-id2}
}
Also, we can use \eqref{full} to express the determinant of a matrix in
terms of traces: 
\equ{
|A| ~=~ \langle\, A^4 \,\rangle ~=~ 
\frac 1{24} \, \Big( 
[A]^4 \,-\, 6 \, [A]^2 \, [A^2] \,+\, 3 \, [A^2]^2 \,+\, 8 \, [A] \, [A^3] \,-\, 6 \, [A^4]
\Big)~. 
}

Finally, if the matrices 
$A, \ldots D$ are formed from five arbitrary vectors 
$p, q, r, s$ and $t$, 
\equ{
A_{\ga a} ~=~ p_\ga q_a~,\quad  
B_{\gb b} ~=~ p_\gb r_b~,\quad  
C_{\gg c} ~=~ p_\gg s_c~,\quad  
D_{\gd d} ~=~ p_\gd t_d~, 
\label{vectormat}
}
then, by the anti--symmetry with respect to the exchange of any two of
the $p_\ga, p_\gb, p_\gg, p_\gd$  inside the bracket expression, we find
$\langle ABCD \rangle = 
\langle \get ABC \rangle = 
\langle \get^2 AB \rangle = 0\,$.

\section{Perturbative expansions}
\label{sc:pertexp}

This Appendix is devoted to some technical details of the perturbative
expansions, that we use in the main part of the text to determine the
interactions of the massive gravity theory in the St\"uckelberg
formulation. The interactions are encoded in the expression
\eqref{accosmo}, where the matrix $\gP$ is defined in
eq.~\eqref{matrixPi}. The first step for the computation of $\Pi$ is
to determine $L$ from eqs.~\eqref{condition}. Since the interactions
of the graviton are determined by an expansion around the Minkowski
background, we consider the infinitesimal general coordinate 
transformation
\equ{
\Id \,+\, \ge \, \get\inv ~=~ \pp[x]{y} \, e \, \get\inv~,
\label{defge}
}
and we expand $L = \sum_n L_n$ in a power series in $\epsilon$ and its
transpose (more accurately, $L_n$ is a sum of monomials of degree $n$,
where each of the monomials is a product of $\epsilon$ and its
transpose). Using this expansion, eqs.~(\ref{condition}) can be
rewritten as as two recursion 
relations 
\equ{
L_{n+1} \,-\, L^T_{n+1} ~=~ L^T_n\eta^{-1}\epsilon^T \,-\, \epsilon \eta^{-1}L_n
~,\qquad
L_{n+1} \,+\, L^T_{n+1} ~=~ \,-\, \sum_{k=1}^n L_k \eta^{-1} L^T_{n-k+1}
~,
}
which, altogether, determine $L$ order by order in the expansion. From
this condition, and from taking $L = \eta$ when the change of
coordinate is trivial ($\epsilon = 0$), we find the recursive solution

\equ{
L_0 ~=~ \eta
~, \quad
L_{n+1} ~=~ \frac{1}{2} \, \Big(
L_n^T \eta^{-1} \epsilon^T \,-\, \epsilon\,\eta^{-1} L_n
\,-\, \sum_{k =1}^n L_k \, \eta^{-1} L_{n-k+1}^T
\Big)~,
\label{recursionrelation}
}
where the last term in parenthesis must be evaluated only for $n > 1
\,$. This determines $L$ uniquely; up to cubic order, the explicit
solution reads 
\equa{\dsp
L ~=~ \, & \dsp \eta \,-\, \frac{1}{2} \, \Big( \epsilon \,-\, \epsilon^T \Big)
\,+\, \frac{1}{8} \, \Big(
3\, \epsilon^2 \,-\, \epsilon \epsilon^T \,-\,\epsilon^T \epsilon \,-\, (\epsilon^{T})^{2}
\Big)
\non \\[2ex]& \dsp
 \,+\, \frac{1}{16} \, \Big(
 \epsilon^2 \,\epsilon^T \,+\, \epsilon\,\epsilon^T\epsilon 
\,+\, \epsilon (\epsilon^{T})^{2} \,+\, \epsilon^T \epsilon^2 
\,-\, \epsilon^T\epsilon\,\epsilon^T \,+\, (\epsilon^{T})^{2} \epsilon 
\,+\, (\epsilon^{T})^{3} \,-\, 5 \, \epsilon^3 
\Big) \,+\, \ldots ~.
\label{L3rd}
}
In this expression the presence of $\eta^{-1}$ between any two
consecutive $\epsilon$ or $\epsilon^T$ is understood.
We can now use this information to determine the matrix
$\gP = \get + \gP_1 + \gP_2 + \gP_3 +\ldots\,$, see \eqref{matrixPi},
where the subscript indicates the order in which the graviton $f$
(before Weyl rescaling) and the pion field $\gp$ appear in this
expression. In fact, in this work we only need the expansion up to
third order: We must evaluate the interaction term \eqref{accosmo} up
to quartic order in the fields, since we are at most interested in
four point interaction vertices. However, since one of the two factors
entering in  \eqref{accosmo} does not have a background part, it is
sufficient to expand $\Pi$ at cubic order in the fluctuations.

From the definition \eqref{matrixPi}, we see that we need to invert
$L$ up to cubic order.  The inversion up to cubic order in $\epsilon$
is obtained in a straightforward manner from (\ref{L3rd}). However,
one has then to realize that $\epsilon$ itself is an expansion series
in terms of the fields which are contained in $f$ and $\pi
\,$. Therefore, we must now expand $\epsilon$ at cubic order in the
physical fields. This is done expanding the definition of $\epsilon
\,$, eq.~(\ref{defge}), where  (in matrix notation) $e = \eta + f$,
while $\pi$ enters in the inverse of $\partial x / \partial y$ as
written in \eqref{matrixpi}. The expansion up to cubic order for
$\epsilon$ reads 
\equ{
\epsilon ~=~
f \,-\, \pi \,-\, \pi\, \eta^{-1}( f\,-\,\pi)
\,+\, \pi \, \eta^{-1} \pi \, \eta^{-1} ( f \,-\, \pi)
\,+\,\ldots
~.
\label{epsiloneq}
}
and, finally, $L^{-1}$ is found to be
\equa{
L^{-1} ~=~\, & \dsp
 \eta^{-1} \,+\, \frac{1}{2} \, \Big(\pi^T \,-\, \pi \Big)
\,+\, \frac{1}{8} \, \Big(3 \, \pi^2 \,-\, (\pi^T )^2 \,-\, \pi\pi^T \,-\,\pi^T\pi \Big) \,+\,
\frac14\,\Big( f \pi^T \,+\, f\pi \,-\, \pi f \,-\, \pi^T f \Big)
\nonumber \\[2ex] & \dsp
\,+\, \frac{1}{16} \, \Big(
(\pi^T)^3 \,-\,5\pi^3 
\,+\, \pi (\pi^T)^2 \,+\, \pi^2\pi^T
\,+\, \pi\, \pi^T\pi \,-\, \pi^T\pi\,\pi^T
\,+\, \pi^T\pi^2 \,+\, (\pi^T)^2\pi \Big)
\nonumber \\[2ex] & \dsp
\,+\, \frac{1}{8} \, \Big(
2 \,\pi^2 f  \,-\, f (\pi^T)^2 
\,-\, \pi f \pi  \,+\, \pi^T f \pi^T
\,-\, \pi f \pi^T
\,-\, f \pi^T \pi
\,+\, \pi^T f \pi
 \,-\, f \pi^2
 \,+\, f \pi \, \pi^T \Big)
\nonumber \\[2ex] 
&\,+\, \frac18 \, \Big( \pi f^2 \,-\, f^2 \pi^T
\,-\,f^2 \pi \,+\, \pi^T f^2 \Big)
\,+\, \ldots~. 
\label{Linv}
}
In both this and the next expressions we have suppressed writing the
matrix $\get\inv$ between consecutive factors. 
Inserting this expression in eq.~\eqref{matrixPi}, we obtain
\equa{
\Pi ~=~ &\dsp
\get \,+\, \frac12 \, \Big( \pi \,+\, \pi^T \Big)
\,+\, \frac{1}{8}\, \Big(3\,\pi\, \pi^T \,-\, \pi^2 \,-\,\pi^T\pi \,-\,(\pi^T)^2 \Big)
\,+\, \frac14\, \Big(f \pi^T \,+\, f\pi \,-\, \pi f \,-\, \pi^T f \Big)
\nonumber \\[2ex]
& \,+\, \frac{1}{16} \,\Big( \pi^3 \,+\, (\pi^T)^3  
\,-\,\pi\,(\pi^T)^2 \,-\, \pi^2\pi^T \,-\, \pi\,\pi^T\pi \,-\, \pi^T\pi\,\pi^T \,+\,
\pi^T\pi^2 \,+\, (\pi^T)^2 \pi \Big)
\nonumber \\[2ex] &
\,+\, \frac{1}{8}\, \Big(
\pi f \pi   \,-\, f (\pi^T)^2 \,+\, \pi^T f \pi^T \,+\, \pi f \pi^T
 \,-\, f \pi^T \pi \,+\, \pi^T f \pi \,-\, f \pi^2  \,+\, f \pi \, \pi^T
\,-\, 2\, \pi\, \pi^T f \Big)
\nonumber \\[2ex] &
\,+\, \frac18 \, \Big( \pi f^2 \,-\, f^2 \pi^T \,-\,f^2 \pi \,+\, \pi^T f^2 \Big)
\,+\, \ldots~.
\label{Pi}
}
which we finally insert in the interaction term~(\ref{accosmo}). This
gives the interactions between the various polarizations of the
graviton for the model we are discussing.

Eq.~\eqref{intergL3} of the main text included the dominant
interaction terms up to quartic order. These leading terms are
obtained from the general expressions just given, by substituting
\eqref{matrixpi} in the expression above, and only keeping terms with at
most a single $f$ or two $A$'s. (All other terms have fewer
derivatives, and so do not control the high energy limit of the
model.) This gives  
\equ{
\gP ~=~ \eta + \gF \,+\, \frac 12 \Big( A \,+\, A^T \Big) 
\,+\, \frac 14 \, \gF (F \,- 2 \, f) \,-\, \frac 14 (F \,- 2 \, f) \gF 
\,+\, \frac 18 \Big( 3 A A^T \,-\, A^2 \,-\, (A^T)^2 \,-\, A^T A \Big)
\non \\[2ex] 
\,+\, \frac 18 \Big( 
F \gF^2 \,-\, \gF^2 F \,+\, F A^T \gF \,-\, \gF A F \,+\, A^T \gF A \,-\, A \gF A^T 
\,+\, 4\, \gF f \gF \,-\, 2\, f \gF^2 \,-\, 2\, \gF^2 f
\Big)~. 
\label{leadingPi}
}
Notice that many possible structures are absent. In particular all 
terms with higher powers of $\gF$ and no other fields have
canceled. The reason for this is that in \eqref{Pi} terms that could
give such terms have coefficients that add up to zero. For example,
when substituting $\gp \ra \gF$ in the combination 
$(3\gp \gp^T -\gp^2 - (\gp^T)^2 -  \gp^T\gp)$ gives zero because $\gF$
is symmetric. This completes the development of the expansions of the
various functions that appear in the main text to the order required there.

\section{Exact expressions}
\label{sc:exact}

In addition to the perturbative expansions presented in the previous
Appendix, it is also possible to derive closed exact expressions. Such
results can be obtained as follows: By multiplying the relations 
\equ{
\pp[x]{y} \,e\, \get\inv ~=~ e' L\inv~, 
\quad \text{and} \quad 
\get\inv e \, \Big(\pp[x]{y}\Big)^T ~=~ (L^T)\inv e'~,
\label{defLeprime}
}
we can obtain an equation for $\get\inv e'$ of which we can take the 
formal square root 
\equ{
e' ~=~ \get \Big[ \get\inv \pp[x]{y} \,e\, \get\inv e 
\Big(\pp[x]{y}\Big)^T \Big]^{1/2}~. 
}
Substituting this back into one of the equations \eqref{defLeprime}, we
find 
\equ{
L ~=~ e \Big(\pp[x]{y}\Big)^T \Big[ 
\get\inv \pp[x]{y} \, e\, \get\inv e \Big(\pp[x]{y} \Big)^T
\Big]^{-1/2}~. 
}
This in turn results in  
\equ{
\gP ~=~ 
\pp[y]{x} \, \Big[ 
\pp[x]{y} \,e\,\get\inv e \Big(\pp[x]{y}\Big)^T \get\inv 
\Big]^{-1/2} \pp[x]{y} \, e 
~. 
}

\end{document}